\documentclass[aps,showpacs,preprint,footinbib,preprintnumbers]{revtex4}

\usepackage{graphicx}
\usepackage{epstopdf}
\usepackage{amsmath,amssymb}
\usepackage{booktabs}
\usepackage{mathrsfs}

\newcommand{\be}{\begin{eqnarray}}
\newcommand{\ee}{\end{eqnarray}}

\newcommand{\ket}{\rangle}
\newcommand{\bra}{\langle}

\newcommand{\vslash}{{v\hspace{-5.4pt}/}}

\usepackage[normalem]{ulem}  
\usepackage[dvips]{color} 
\renewcommand\sout{\bgroup \color{red} \ULdepth=-.5ex \ULset}

\begin{document}

\title{Exotic baryons from a heavy meson and a nucleon \\ -- Negative parity states --}
\
\author{Yasuhiro Yamaguchi$^1$}
\author{Shunsuke Ohkoda$^1$}
\author{Shigehiro Yasui$^2$}
\author{Atsushi Hosaka$^1$}
\affiliation{$^1$Research Center for Nuclear Physics (RCNP), 
Osaka University, Ibaraki, Osaka, 567-0047, Japan}
\affiliation{$^2$KEK Theory Center, Institute of Particle and Nuclear
Studies, High Energy Accelerator Research Organization, 1-1, Oho,
Ibaraki, 305-0801, Japan}

\date{\today}


\begin{abstract}
We study heavy baryons with an exotic flavor quantum number 
formed by a  heavy  meson and a nucleon ($\bar DN$ and $BN$)
through a long range one pion exchange interaction.  
The bound state found previously in the $(I,J^P) =(0,1/2^-)$ channel survives 
when short range interaction is included.  
In addition, we find a resonant state with $(I,J^P)=(0,3/2^-)$ 
as a Feshbach resonance 
predominated by a heavy  vector meson and a nucleon ($\bar D^*N$ and $B^*N$).  
We find that these exotic states exist for the charm and heavier flavor region.
\end{abstract}
\maketitle

\section{Introduction}

Hadron physics has opened a renewed interest in multi-hadron systems.  
The most familiar example is the atomic nucleus which 
is the bound state of protons and neutrons.  
However, interestingly enough, we do not have yet clear evidences for analogous 
systems of baryon number one or zero.  
Yet a well-known candidate is $\Lambda(1405)$ 
which is considered to be a quasi-bound state
of $\bar K N$ and $\pi \Sigma$ \cite{Oset:1997it,Hyodo:2002pk,Hyodo:2003qa,Hyodo:2007jq,Jido:2003cb}.  
The isoscalar meson which has been established as a resonance 
in pseudoscalar meson scatterings
may also be a quasi-bound state of the two mesons \cite{Hyodo:2010jp,PhysRevD.56.3057,Oller:1997ti,PhysRevD.59.074001}.    
Such hadronic composites, or molecular states, 
are dynamically generated via  hadron-hadron interactions, 
and are expected to appear in various mesonic and baryonic systems.
 The hadron composites are important also for study of the hadron dynamics in nuclear matter \cite{Jido:2002yb,Mizutani:2006vq,Tolos:2007vh,Dote:2008in,Dote:2008hw,Lutz:2009ff}.
 
In the constituent quark model, the generation of hadronic composites can be 
understood as formation of clusters in multi-quark systems.
Because a typical excitation energy of hadron resonances amounts to 
several hundred MeV and is enough to create a $\bar q q$ pair of constituent quarks, 
a multi-quark component naturally appears in the resonance states in addition 
to the minimal configuration of   $\bar qq$ or $qqq$ with an orbital excitation.  
Such a multi-quark configuration may arrange itself into a set of color-singlet clusters, 
namely a set of hadrons.  
This serves a quark model picture of hadronic composites.  

%

Recently, a novel structure has been suggested  by one of the present
authors in manifestly exotic channels with one (anti) heavy quark,
 for example, $\bar D N$ whose minimum quark content is 
$uudd\bar c$~\cite{yasui}.
This is a charm analogue of the pentaquark $\Theta^+ \sim uudd \bar s$
~\cite{Diakonov:1997mm,Nakano:2003qx}.
So far the charmed pentaquarks in various forms have been discussed by many authors \cite{Gignoux:1987cn,Lipkin87,Riska:1992qd,Oh:1994yv,Oh:1997tp,Stewart:2004pd,Sarac:2005fn,Cohen:2005bx,Lee:2007tn,Lee:2009rt,Wu:2010jy}.
To obtain a bound $\bar DN$ system composed of a $\bar{D}$ meson and a nucleon, 
the one pion exchange interaction was found to be crucially important.
As emphasized in Ref.~\cite{yasui}, in particular, the tensor force yields strong attraction through 
the mixing of an $S$-wave state of $\bar DN$ and a $D$-wave state of $\bar D^* N$.  
The mixing effect is more important for heavier flavor sectors, where
pseudoscalar and vector mesons are more degenerate. 
Indeed the mass splittings of $\bar D \bar D^*$ and of $B B^*$ 
are about 140 and 46 MeV, respectively, 
and are much smaller than the one of the strangeness sector ($KK^*$ mass difference
$ \sim 400$ MeV).  

Such a mixing mechanism of $S$- and $D$-waves (more generally 
mixing of $L$- and ($L \pm2$)-waves, with $L$ being an orbital angular
momentum) 
has been known to be very important for the deuteron binding, while 
its relevance has been reexamined  
for other nuclear systems rather recently~\cite{Pieper:2001mp,Ikeda:2010aq}.  
In QCD, spontaneous breaking of chiral symmetry is underlying; 
the light pseudoscalar mesons, pions, are generated as 
the Nambu-Goldstone modes, with  strong coupling to hadrons \cite{Nambu:1961tp,Nambu:1961fr}.  
It is emphasized that the pseudoscalar nature of the pion 
necessarily leads to the Yukawa coupling of the form
$\vec{s} \cdot \vec{q}$, where $\vec{s}$ is the spin operator for 
the particles coupling to the pion and $\vec{q}$ is the 
momentum of the pion.  
Then this coupling provides the tensor force in two-body systems.  
It is noticeable that such one pion exchange interaction can lead
also to stable exotic states of two heavy meson
systems as pointed out in Refs.~\cite{Cohen:2005bx,He:2010zq,Manohar:1992nd,Tornqvist:1993ng}.
In the present study, we investigate the exotic state of a heavy meson
and a nucleon, namely $\bar{D}N$ and $BN$.

In this paper, we examine the system of $PN$ and $P^*N$ with 
the inclusion of short range interactions.  
Here and in what follows we introduce the notation 
$P (= \bar D, B)$ for a heavy  pseudoscalar meson and 
$P^* (= \bar D^*, B^*)$ for a heavy  vector meson.  
We study  not only bound states but also resonances.  
We concentrate our analysis on the low lying states in which $S$-wave component is included.
Furthermore, we present our analysis only in isospin singlet channels because 
we find neither bound state nor resonant state in isospin triplet
channels.

In section II, 
we briefly describe the interactions between $PN$ and $P^*N$ 
based on the heavy quark symmetry.  
It has been known that the heavy quark symmetry plays an important role
for charm and bottom quarks, not only in the dynamics of
quarks~\cite{wise,Isgur:1991wq,Manohar:2000dt}, but also in the dynamics
of heavy
hadrons~\cite{Manohar:2000dt,Burdman:1992gh,Wise:1992hn,Yan:1992gz,Nowak:1992um,Bardeen:1993ae}.
Therefore, in the present study, we employ the interaction from heavy quark symmetry.
For comparison, we also investigate the interaction derived from flavor SU(4) symmetry~\cite{Lutz:2003jw,Hofmann:2005sw,Gamermann:2006nm,Haidenbauer:2007jq}.
In this paper, we include not only the pion exchange but also vector meson exchange interactions 
to show the dominant role of the pion exchange interaction.  
We discuss bound states and scattering states in sections III and IV, respectively.  
We confirm that a bound state exists for isospin, spin and parity
$(I,J^P) =(0,1/2^-)$.
For resonance, we find a new  state in the $(I,J^P) = (0,3/2^-)$ channel having  
a narrow decay width as a Feshbach resonance predominated by  $P^* N$.  
In section V, we discuss  flavor dependence 
of the present results by varying the mass of the heavy mesons continuously, 
and show that the existence of the above exotic states is the feature 
of heavy flavors.   
Final section is devoted to summary and discussions. 

\section{Interactions}

The systems we are interested in have a manifestly exotic flavor
 structure of $ qqqq \bar{Q}$ in its minimal quark content, 
which is a heavy quark analogue of the pentaquark 
$\Theta^+ \sim  qqqq \bar{s}$.  
Here $Q$ and $q$ denote heavy and light ($u, d$) quarks.  
We investigate  whether these exotic baryons are formed as (quasi) bound states of a $\bar Q q$ 
meson (denoted by $P$ or $P^*$ following introduction) 
and a $ qqq$ nucleon  (denoted by $N$).  
For this picture to work well, 
these two hadronic ingredients should be sufficiently apart 
and keep their identities.  
Possible effects of internal structure is then expressed by form factors.  
The two-body states of a pseudoscalar (vector) meson and a nucleon are classified by their isospin $I$, total spin $J$ and parity $P$, 
or orbital angular momentum $L$, where $P = (-1)^{L+1}$.  
For a given $J^P$, there are three or four coupled channels as summarized 
in Table.~\ref{table_qnumbers}, where low lying components  including 
$S$-states are shown.  

\begin{table}[tbp]
\centering
\caption{\label{table_qnumbers} \small Various coupled channels for a 
given quantum number $J^P$ for negative parity $P = -1$.  }
\vspace*{0.5cm}
{\small 
\begin{tabular}{ c  | c c c c}
\hline
$J^P$ &  \multicolumn{4}{c }{channels} \\
\hline
$1/2^-$ &  $PN(^2S_{1/2})$ & $P^*N(^2S_{1/2})$ & 
     $P^*N(^4D_{1/2})$ & \\
$3/2^-$ &  $PN(^2D_{3/2})$ &  $P^*N(^4S_{3/2})$ & 
     $P^*N(^4D_{3/2})$ & $P^*N(^2D_{3/2})$ \\
\hline
\end{tabular}
}
\end{table}

To obtain interactions for heavy mesons and nucleons, we employ Lagrangians 
satisfying  heavy quark symmetry and chiral symmetry \cite{Casalbuoni}.  
They are well-known and given as
\be
{\cal L}_{\pi HH} &=&   ig_\pi \mbox{Tr} \left[
H_b\gamma_\mu\gamma_5 A^\mu_{ba}\bar{H}_a \right]   \, , 
\label{LpiHH}\\
{\cal L}_{v HH} &=&  -i\beta\mbox{Tr} \left[ H_b
v^\mu(\rho_\mu)_{ba}\bar{H}_a \right]
+i\lambda\mbox{Tr} \left[
H_b\sigma^{\mu\nu}F_{\mu\nu}(\rho)_{ba}\bar{H}_a \right] \, , 
\label{LvHH}
\ee
where the subscripts $\pi$ and $v$ are for the pion 
and vector meson ($\rho$ and $\omega$) interactions, and 
$v^{\mu}$ is the four-velocity of a heavy quark.
In Eqs.~\eqref{LpiHH} and \eqref{LvHH}, the heavy meson fields of $\bar Q q$ are parametrized by 
the heavy pseudoscalar and vector mesons, 
\be
H_a   &=& \frac{1+\vslash}{2}\left[P^\ast_{a\,\mu}\gamma^\mu-P_a\gamma_5\right] \, ,  \\
\bar H_a  &=& \gamma_0 H^\dagger_a \gamma_0 \, ,
\ee
where the subscripts $a, b$ are for light flavors ($u, d$).  
 The pseudoscalar and vector fields are normalized as 
\be
\bra 0| P |P(p_\mu)\ket &=& \sqrt{p^0}\, ,  \\
\bra 0| P^*_\mu |P^*(p_\mu, \lambda)\ket &=& 
\epsilon(\lambda)_\mu \sqrt{p^0}\, , 
\ee
where $\epsilon(\lambda)_{\mu}$ is the polarization vector of $P^*$ with 
polarization $\lambda$.  
The axial and vector currents of light flavors are given by 
\be
A^{\mu} & =\frac{1}{2}(\xi^\dagger\partial^\mu \xi-\xi\partial^\mu
\xi^\dagger) 
\, , \label{Acurrent}\\
V^{\mu} & =\frac{1}{2}(\xi^\dagger\partial^\mu \xi+\xi\partial^\mu
\xi^\dagger) 
\, , \label{Vcurrent}
\ee
where $\xi=\exp(i\hat{\pi}/f_{\pi})$, $f_{\pi} = 132$ MeV 
is the pion decay constant and we define the pion field by
\begin{align}
   \hat{\pi}  &= 
\begin{pmatrix}
 \frac{\pi^0}{\sqrt{2}} & \pi^+ \\
 \pi^- & - \frac{\pi^0}{\sqrt{2}} 
\end{pmatrix} .
\end{align}
Finally the vector ($\rho$ and $\omega$) meson field and its field tensor are defined by 
\begin{eqnarray}
\rho_{\mu} &=& i\frac{g_V}{\sqrt{2}}\hat{\rho}_{\mu} \, , \\
   \hat{\rho}_{\mu} &=&
\begin{pmatrix}
 \frac{\rho^0}{\sqrt{2}} + \frac{\omega}{\sqrt{2}} & \rho^+ \\
 \rho^- & - \frac{\rho^0}{\sqrt{2}} +\frac{\omega}{\sqrt{2}}
\end{pmatrix}_{\mu} \, , \\
 F_{\mu \nu}(\rho) &=& \partial_{\mu}\rho_{\nu}
  -\partial_{\nu}\rho_{\mu} +[ \rho_{\mu} ,\rho_{\nu}] \, ,
\end{eqnarray}
%
where $g_V$ is the gauge coupling constant of the hidden local symmetry \cite{Bando:1987br}.

The coupling constant $g_\pi$ for $\pi PP^\ast$ is fixed from the strong
decay of $D^\ast \rightarrow D\pi$~\cite{PDG}.
The coupling constants $\beta$ and $\lambda$ are determined by the
radiative decays of $D^\ast$ and semileptonic decays of $B$ with the vector
meson dominance~\cite{Isola}. 
The resulting values are given in Table.~\ref{table_Nconstants}.

From Eq.~(\ref{LpiHH}) we obtain  $\pi PP^*$ and $\pi P^* P^*$ 
vertices as
\be
{\cal L}_{\pi PP^*} &=& 
2 \frac{g_\pi}{f_\pi}(P^\dagger_a P^\ast_{b\,\mu}+P^{\ast\,\dagger}_{a\,\mu}P_b)\partial^\mu\hat{\pi}_{ab}
 \, , 
\label{LpiPP*} \\
{\cal L}_{\pi P^*P^*} &=& 
2 i \frac{g_\pi}{f_\pi}\epsilon^{\mu \nu \alpha \beta} v_{\mu}
P^{\ast\,\dagger}_{a\,\beta}P^\ast_{b\,\nu}\partial_{\alpha}
\hat{\pi}_{ab}
 \, . 
\label{LpiP*P*}
\ee
We note that there is no $\pi PP$ vertex due to parity invariance.  
Similarly, from Eq.~(\ref{LvHH}), we
derive the vector meson vertices as 
\be
{\cal L}_{vPP} &=&
-\sqrt{2}\beta g_V P_b P^{\dagger}_a v \cdot \hat{\rho}_{ba}
 \, , 
\label{LvPP} \\
{\cal L}_{vPP^*} &=& 
-2\sqrt{2}\lambda g_V v_{\mu}\epsilon^{\mu \nu \alpha \beta}
\left(P^\dagger_a
P^\ast_{b\,\beta}-P^{\ast\,\dagger}_{a\,\beta}P_b\right)
\partial_\nu(\hat{\rho}_\alpha)_{ba}
\, ,
\label{LvPP*} \\
{\cal L}_{vP^*P^*} &=&
\sqrt{2} \beta g_V P^*_b P^{*\dagger}_a v \cdot \hat{\rho}_{ba}\nonumber \\
&&+i2 \sqrt{2}\lambda g_V
P^{\ast\,\dagger}_{a\,\mu}P^\ast_{b\,\nu}
(\partial^\mu(\hat{\rho}^\nu)_{ba}-\partial^\nu(\hat{\rho}^\mu)_{ba}) \, .
\label{LvP*P*}
\ee

The interaction Lagrangians for a meson and nucleons are given by the standard form, 
\be
{\cal L}_{\pi NN} &=& \sqrt{2} ig_{\pi
NN}\bar{N}\gamma_5 \hat{\pi} N \, , \label{LpiNN} \\
{\cal L}_{vNN} &=& \sqrt{2} g_{vNN}\left[\bar{N} \gamma_\mu \hat{\rho}^\mu N 
+\frac{\kappa}{2m_N}\bar{N} \sigma_{\mu\nu} \partial^\nu \hat{\rho}^\mu
N \right] \, , \label{LvNN}
\ee
where $N =(p,n)^T$ is the nucleon field.
In the vector meson interaction there are vector (Dirac) and tensor (Pauli) 
terms.  
For $\rho$ meson the tensor term dominates, while for $\omega$ it is negligible.  
The coupling constants associated to the nucleon are taken from Ref.~\cite{machleidt} 
as summarized also in Table.~\ref{table_Nconstants}.

\begin{table}[tbp]
\centering
\caption{\label{table_Nconstants} Masses and coupling constants of mesons.}
\vspace*{0.5cm}
{\small 
\begin{tabular}{ c  c c c c c c}\hline
  & $m_{\alpha}\,\,[\mbox{MeV}]$ & 
 $g_\pi$ & $\beta$ & $\lambda\,\,[\mbox{GeV}^{-1}]$&  $g^2_{\alpha NN}/4\pi$ & $\kappa$  \\
\hline
$\pi$ & 137.27 & 0.59 & --- & ---  &  13.6 & ---  \\
$\rho$ & 769.9 & --- & 0.9 & 0.56 & 0.84 & 6.1 \\
$\omega$ & 781.94 & --- & 0.9 & 0.56 & 20.0 & 0.0 \\ \hline
\end{tabular}
}
\end{table}


To parametrize internal structure of hadrons, we introduce form factors associated 
with finite size of the mesons and nucleons.  
We adopt a dipole form factor at each vertex :
\begin{align}
 F_\alpha (\Lambda,\vec{q}\,)&=\frac{\Lambda^2-m^2_{\alpha}}{\Lambda^2 +
 |\vec{q}\,|^2} \, ,
 \label{eq:cutoff}
\end{align}
where $m_\alpha$ and $\vec{q}$ are the mass and three-momentum of the
incoming meson $\alpha$ ($=\pi,\rho,\omega$). The cutoff parameter
 for the meson-nucleon vertex $\Lambda = \Lambda_N$ is determined such that the
resulting $NN$ potential reproduces the binding energy of the
deuteron. For the derivation of the potential, see the
discussion below and also Appendix.
When the $NN$ potential is constructed only by $\pi$ exchange,
$\Lambda_N=830$ MeV, while when the potential is constructed by
$\pi,\rho,\omega$ exchanges, $\Lambda_N=846$ MeV.
To test the validity of the potentials, we have computed low energy
properties of the deuteron and $NN$ scattering. The results are given in
Table.~\ref{table_deuteron}, for the $\pi$ exchange and
$\pi,\rho,\omega$ exchange potentials.
Another cutoff parameter $\Lambda_P$ for the meson-meson vertex is
determined by the ratio of the size of the pseudoscalar meson $P$ as
given in Ref.~\cite{yasui}, $\Lambda_D=1.35\Lambda_N$ and
$\Lambda_B=1.29\Lambda_N$.
We have adopted the same value of the cutoff for the vertices including
vector meson $P^\ast$.

We have also constructed the interaction from the flavor SU(4) 
symmetry for comparison with that of the heavy quark symmetry.
However, it turns out that the coupling strength of the
$\pi PP^*$ vertex is smaller than the one determined from the experimental
decay rate of $D^* \rightarrow D\pi$.
We have found that there is 
no bound state when such a small coupling strength is used.
In this work we use the potential derived from the heavy quark symmetry
with the coupling strength determined by  the experimental decay
width of $D^* \rightarrow D\pi$.

Having all the above interaction Lagrangians for the Yukawa vertices, 
we obtain  potentials for various channels which
are summarized in Appendix.  
In doing so, we employ the static approximation where 
the energy transfer can be ignored.  
This is a good approximation when $P^*$ does not decay into $P \pi$ 
in the heavy quark limit and at low energies below the threshold of 
pion productions.  
The total Hamiltonian is then given as the sum of the kinetic energies 
and the potential for coupled channels as shown explicitly in Appendix.
We solve the coupled Schr\"odinger equations for the $PN$ and $P^*N$ system in their center of
mass frame by using the numerical method developed in
Ref.~\cite{johnson}. The total energy $E$ is measured from the $PN$ threshold. 

We have tested our method for the deuteron case where we can compare 
our results with the known results.  
As summarized in Table.~\ref{table_deuteron}, we point out that the $\pi$ 
exchange potential  reproduces the deuteron properties well, such as the binding energy $E_{B}$, the $D$-wave probability $P_{D}$, the mean square radius $\langle r^{2} \rangle^{1/2}$, as well  as the scattering length $a$ and the effective range $r_{e}$ in the $^{3}S_{1}$ and $I=0$ channel in the $NN$ scattering.  
The results with the $\pi,\rho,\omega$ exchange potential are very
similar to those of the $\pi$ exchange potential.
This indicates that the $\rho$ and $\omega$ exchanges play only a minor role 
for low energy properties as expected, because 
the deuteron is a loosely bound state and a rather extended object.  

\section{Bound states}
 

\begin{table}[tbp]
  \caption{\label{table_deuteron} Low energy properties of the $NN$
 system. Results for the $\pi$ and $\pi\rho\,\omega$ potentials are compared.}
 \begin{center}
  \begin{tabular}{lcccccc}\hline
Potential & $\Lambda_N$ [MeV] &  $E_B$ [MeV] & 
 $P_D [\%]$  &  $\bra r^2\ket^{1/2}$ [fm] & $a$ [fm] & $r_e$ [fm] \\ \hline 
$\pi$ & 830 & 2.22 & 5.4 & 3.7 & 5.27 & 1.50 \\
$\pi \rho \, \omega$ & 846 & 2.22 & 5.3 & 3.7 & 5.23 & 1.49 \\ \hline 
 \end{tabular}
 \end{center}
\end{table}


In this section let us study low lying  states for the $PN$-$P^* N$ system. 
We find a bound state in the $(I,J^P) = (0,1/2^-)$
channel as discussed in Ref.~\cite{yasui}. The results of $\pi$ exchange
potential are almost the same as those of $\pi, \rho, \omega$ exchange
potential as seen in the deuteron case.
The binding energies $E_B =|E|$ and the root mean square radii $\langle r^{2} \rangle^{1/2}$ for
$\bar{D}N$ and $BN$ states are shown 
in Table.~\ref{table_result}.  
As emphasized in Ref.~\cite{yasui},  the tensor force causing 
mixing between $\bar DN$ and $\bar D^* N$ states of different angular momenta 
by $\Delta L = 2$ yields strong attraction.  
This is particularly so for heavier quark sector, where $P$ and $P^*$ mesons 
degenerate.  
Therefore, the $BN$ state is more bound than the $\bar{D}N$ state.
In fact, the binding energies of $BN$ states are about ten times larger
than those of $\bar{D}N$ states.
The mass dependence of the binding energies will be discussed more in
detail in section ~\ref{section5}.

\begin{table}[tbp]
 \begin{center}
  \caption{\label{table_result} Binding energies, root mean
  square radii and cutoff parameters of heavy mesons. Results for the $\pi$ and $\pi\rho\,\omega$ potentials are compared.}
  \label{kekka}
  \begin{tabular}{lcccc}\hline
 & $\bar{D}N(\pi)$ &  $\bar{D}N(\pi \rho \, \omega)$ & 
 $BN(\pi)$ &  $BN(\pi \rho \, \omega)$ \\ \hline 
$E_B$ [MeV] & 1.60 & 2.14 & 19.50 & 23.04 \\
$\bra r^2\ket ^{1/2}$ [fm] & 3.5 & 3.2 & 1.3 & 1.2 \\
$\Lambda_P$ [MeV] & 1121 & 1142 & 1070 & 1091 \\
 \hline 
 \end{tabular}
 \end{center}
\end{table}

\begin{figure}[btp]
\begin{minipage}{0.45\hsize}
  \begin{center}
   \mbox{ a) $\bar{D}N$}
    
   \scalebox{.65}{\includegraphics{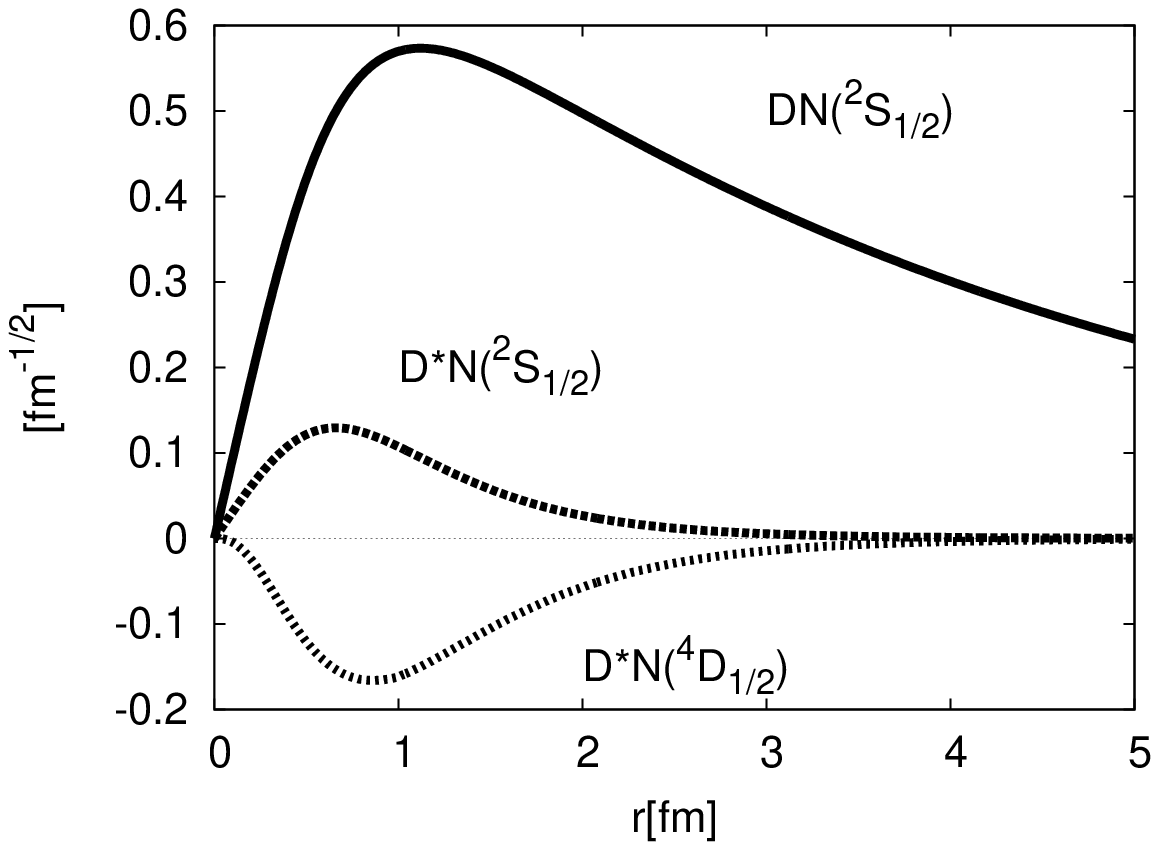}}
  \end{center}
 \end{minipage}
 \begin{minipage}{0.45\hsize}
  \begin{center}
   \mbox{ b) $BN$}
   
   \scalebox{.65}{\includegraphics{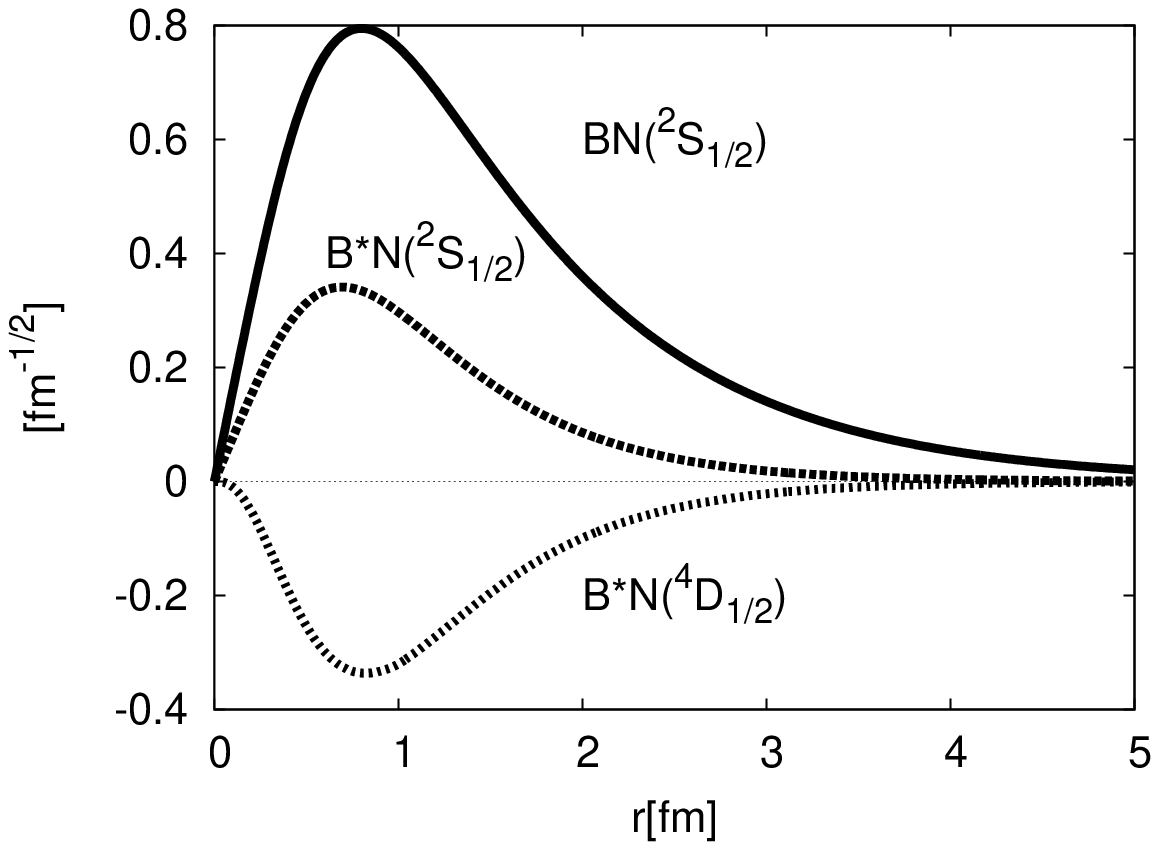}}
  \end{center}
 \end{minipage} 
\caption{\label{figure_wave}The wave functions of the $\bar{D}N$ and
 $BN$ bound states with $(I,J^P)=(0,1/2^-)$ when the $\pi\rho\,\omega$
potential is used.}
\label{fig2}
\end{figure}

The corresponding root mean square radii are over 3 fm for the charm
and over 1 fm for the bottom baryons, respectively.  
Both radii are larger than typical hadron size of order 1 fm, 
justifying the hadronic composite structure of the present states.  
To complete our presentations, in Fig.~\ref{fig2}, we also show bound state 
wave functions when the $\pi\rho\,\omega$ potential is used.
Finally we note that there is no bound state in $J^P = 3/2^-$ as shown
in Ref.~\cite{yasui}  

\section{Scattering states and resonance}

\begin{figure}[h]
 \begin{center}
\begin{tabular}{cc}
{\includegraphics[width=80mm,clip]{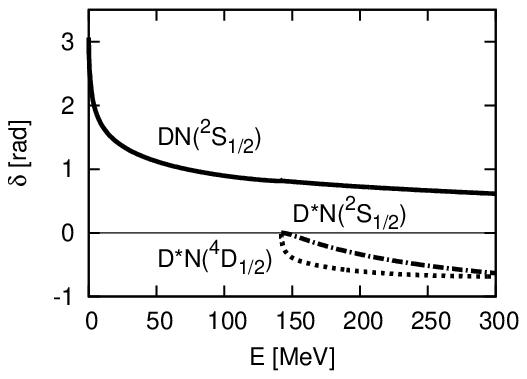}} &
{\includegraphics[width=80mm,clip]{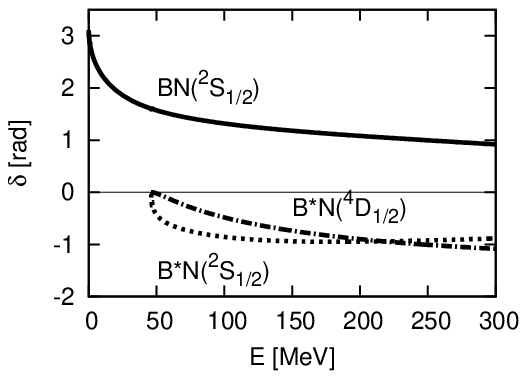}}
\\
\end{tabular}  
\caption{Phase shift of the $\bar{D}N$ and $BN$ scattering state with
  $(I,J^P)=(0,1/2^-)$  when the $\pi\rho\,\omega$ potential is used.}
\label{fig3}
 \end{center}
\end{figure}

Let us now turn to
the scattering region 
above the $PN$ threshold.  
First, we see the scattering state in the $(I,J^P)=(0,1/2^-)$ channel.
In Fig.~\ref{fig3}, we present 
the phase shifts $\delta$
of $PN(^2S_{1/2})$, $P^\ast N(^2S_{1/2})$ and $P^\ast N(^4D_{1/2})$
channels when the $\pi\rho\,\omega$ potential is used.
Each phase shift is plotted as a function of the scattering energy $E$
in the center of mass system.
The $P N$ phase shift starts at $\delta = \pi$ because of the 
presence of the bound state discussed in the previous section.  
Otherwise the energy dependence of all the phase shifts is rather smooth.  
We summarize the scattering lengths and the effective ranges in
Table~\ref{length}.
We have checked the relation $E_B = \frac{1}{2\mu a^2}$ ($\mu$ is reduced mass)
holds with good accuracy within a few percents.
We also note that the properties of $\bar{D}N$ system are rather similar 
to those of the $NN$ system.

\begin{figure}[h]
 \begin{center}
\begin{tabular}{c}
{\includegraphics[width=80mm,clip]{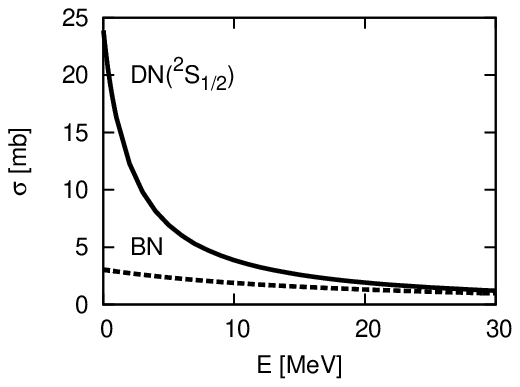}}
\end{tabular}  
\caption{Total cross section of the $\bar{D}N$ and $BN$ scattering state
  with $(I,J^P)=(0,1/2^-)$  when the $\pi\rho\,\omega$ potential is used.}
\label{fig5}
 \end{center}
\end{figure}

We show  the  total cross sections of $\bar{D}N$ and $BN$ 
scattering when the $\pi\rho\,\omega$
potential is used in Fig.~\ref{fig5}.
They start from the maximum value at the threshold and decrease monotonically.
For shallower bound state (for $\bar DN$ system), the peak value 
at the threshold 
is larger, due to the presence of the bound state near the threshold.  
In the limit the binding energy $E_B \to 0$, the peak diverges as
explained by the zero-energy resonance.  

\begin{table}[h]
\caption{The scattering length $a$ and effective range $r_e$ with $(I,J^P)=(0,1/2^-)$.}
\label{length}
 \begin{center}
  \begin{tabular}{lcccc}\hline
 &$\bar{D}N(\pi)$&$\bar{D}N(\pi\rho\,\omega)$&$BN(\pi)$&$BN(\pi\rho\,\omega)$ \\
\hline
$a$ [fm]&4.36&4.38&1.61&1.56 \\
$r_{e}$ [fm]&1.04&1.05&0.71&0.68 \\
\hline
  \end{tabular}
 \end{center}
\end{table}


\begin{figure}[h]
 \begin{center}
\begin{tabular}{cc}
{\includegraphics[width=80mm,clip]{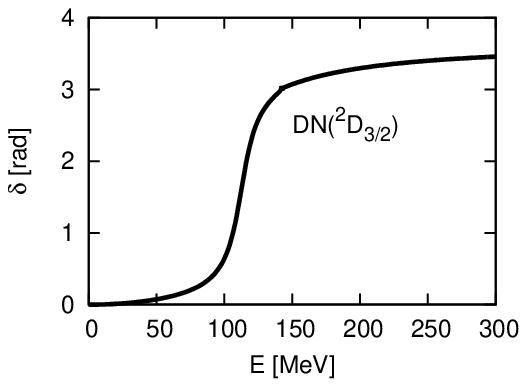}} &
{\includegraphics[width=80mm,clip]{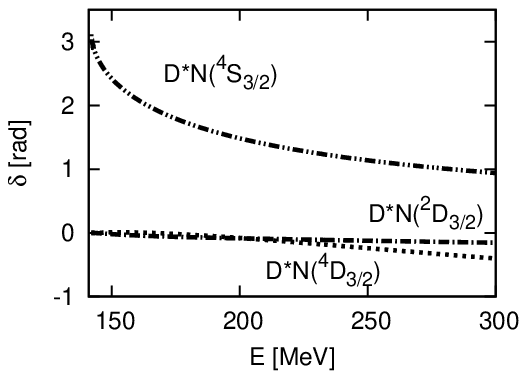}} \\[2mm]
%
{\includegraphics[width=80mm,clip]{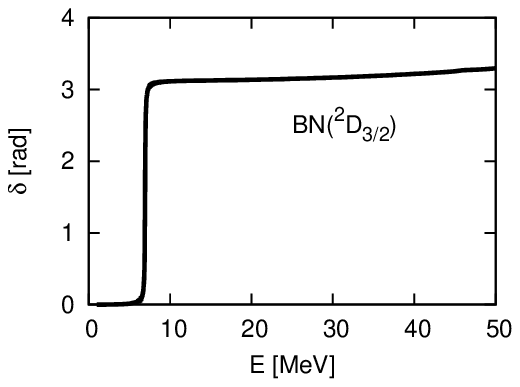}} &
{\includegraphics[width=80mm,clip]{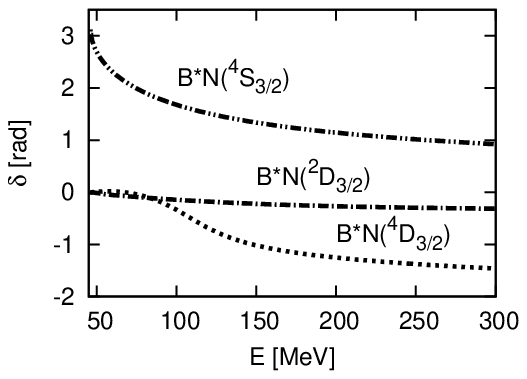}}
\\
\end{tabular}  
\caption{Phase shift of the $\bar{D}N$ and $BN$ scattering state with
  $(I,J^P)=(0,3/2^-)$  when the $\pi\rho\,\omega$ potential is used.}
\label{fig6}
 \end{center}
\end{figure}

In the $(I,J^P)=(0,3/2^-)$ channel, 
we find an interesting structure; 
the phase shifts shown in Fig.~\ref{fig6} indicate a resonance 
at the scattering energy $E_{re}= 113.19$ MeV for $\bar DN$ and 
at $E_{re}= 6.93$ MeV for  $BN$, as the phase shifts cross $\pi/2$.  
The mechanism of the resonance can be understood by looking at the 
phase shifts in other channels; in particular, we find that 
those of $\bar D^* N(^4S_{3/2})$ and of $B^*N(^4S_{3/2})$ start 
from $\delta = \pi$, indicating the presence of a bound state
in these channels. 
Indeed, we checked that, when the $PN(^2D_{3/2})$ channel is ignored
and only the $P^\ast N(^2D_{3/2})$, $P^\ast N(^4D_{3/2})$ and $P^\ast
N(^4S_{3/2})$ channels are considered, 
there are bound states at 
$E_B = 11.50$ MeV from the $\bar D^* N$ threshold and at
$E_B = 21.67$ MeV from the $B^* N$ threshold.  
Therefore, these resonances are the  Feshbach resonances.  

We can also extract the width of the resonances from the slope of 
the phase shift at $\pi/2$; 
\be
\Gamma =
\frac{2}{d\delta/dE|}_{E=E_{re}}
  \, ,
\ee
for the phase shift $\delta$ in the partial wave, namely $PN(^{2}D_{3/2})$.
The results are listed in Table~\ref{table4}.
These values are very small in particular for the $BN$ system.
There are two reasons for that; 
one is that the resonance energy is close to the threshold of the open  
channel of $BN$, 
and the other is that 
the decay occurs in the $D$-wave.  
The latter effect is important as the decay width is proportional to 
the fifth power ($2L +1$, $L = 2$) of the decaying momentum and 
works as a suppression factor near the threshold.  
The physics behind is the centrifugal barrier for a partial wave with 
a finite angular momentum.  

\begin{figure}[h]
 \begin{center}
\begin{tabular}{cc}
{\includegraphics[width=80mm,clip]{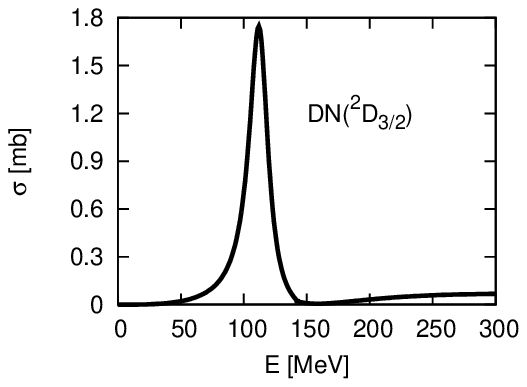}} &
{\includegraphics[width=80mm,clip]{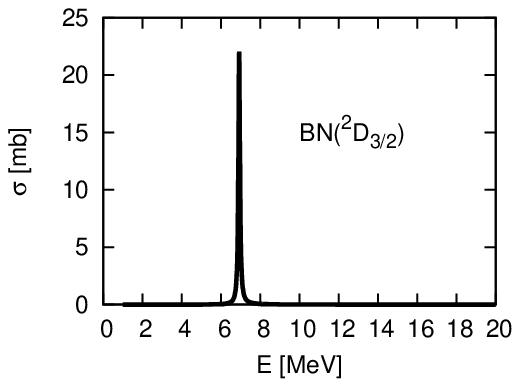}}
\\
\end{tabular}  
\caption{Total cross section of the $\bar{D}N$ and $BN$ scattering state with
  $(I,J^P)=(0,3/2^-)$  when the $\pi\rho\,\omega$ potential is used.}
\label{fig7}
 \end{center}
\end{figure}

In Fig.~\ref{fig7}, we plot the cross section in the $(I,J^P)=(0,3/2^-)$
channel.
Because there is a resonant state at $E_{re}= 113$ MeV in the
$\bar{D}N$ scattering  and at $E_{re}= 7$ MeV in the $BN$ scattering, the cross
section becomes maximum at each resonance energy.

\begin{table}[h]
\caption{The resonance energy and decay width for $(I,J^P)=(0,3/2^-)$.}
\label{table4}
 \begin{center}
  \begin{tabular}{lcccc}\hline 
 &$\bar{D}N(\pi)$&$\bar{D}N(\pi\rho\,\omega)$&$BN(\pi)$&$BN(\pi\rho\,\omega)$ \\
\hline
$E_{re}$ [MeV]&113.51&113.19&8.41&6.93 \\
$\Gamma$ [MeV]&19.43&17.72&0.16& $0.0946$ \\
\hline
  \end{tabular}
 \end{center}
\end{table}

\clearpage
\section{Flavor dependence of the bound and resonant state} \label{section5}

\begin{figure}[h]
 \begin{center}
\begin{tabular}{cc}
\begin{minipage}[cbt]{0.5\textwidth}
{\includegraphics[width=80mm,clip]{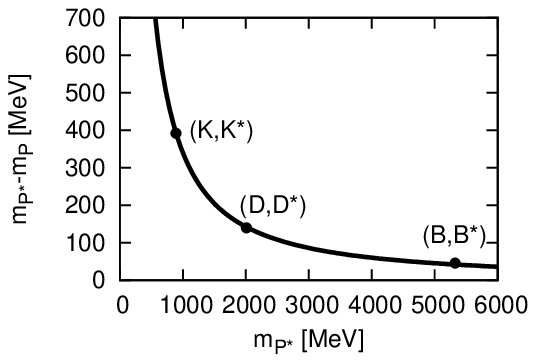}}
\caption{The mass splitting $m_{P^*}-m_P $ as a function of the heavy
 vector meson mass $m_{P^*}$.}
\label{fitmass} 
\end{minipage}
 &\quad\quad
\begin{minipage}[cbt]{0.5\textwidth}
 {\includegraphics[width=80mm,clip]{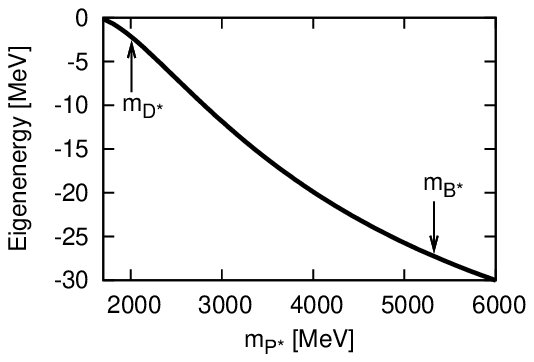}}
\caption{Eigenenergy for $(I,J^P)=(0,1/2^-)$ 
as a function of heavy vector meson mass $m_{P^*}$.  }
\label{eigenmass}
\end{minipage}
 \\
\end{tabular}
 \end{center}
\end{figure}

In the previous sections, we have seen that the bound and 
resonant states in the bottom sector receive more attractions 
than in the charm sector.  
In other words, the existence of the present exotic baryons 
is a unique feature of heavier mesons.  
To see this more quantitatively, let us vary the mass of 
the heavy mesons continuously and see how the bound 
and resonant state properties change. 
Let us make an interpolation of heavy meson masses, by regarding the mass 
of the $P$ meson as a function of the mass of the $P^*$ meson.  
Given the experimental values of 
$(m_K, m_{K^*})$, $(m_{\bar D}, m_{\bar D^*})$ and 
$(m_B, m_{B^*})$, we find the parametrization 
$m_{P^*} - m_{P} = 1.9\times 10^{6}/m_{P^*}^{1.25}$ in  units of MeV to optimally 
interpolate the masses at the three flavor points, as shown 
in Fig.~\ref{fitmass}.
In the heavy quark mass limit, the mass difference 
$m_{P^*} - m_{P}$ is expected to be proportional to the inverse 
mass of the heavy quark and so to the mass of the vector meson.  
The deviation of the power 1.25 from unity is due to some finite mass 
corrections.  
Here, the details of the functional form is not important, but only 
an interpolation at a qualitative level is enough.  

Having this parametrization, we  perform calculations 
for the bound and resonant state properties.  
In Fig.~\ref{eigenmass}, the eigenenergy of the bound state is plotted
as a function of $m_{P^*}$.  
As expected, the binding energy increases as the mass of the heavy 
meson increases.
What is then interesting is that the bound state disappears at 
$m_{P^*} \sim 1700$ MeV, which is about 300 MeV 
smaller than the mass of $D^*$.  
Hence, that the masses of the mesons are heavy is crucial for the presence
of the exotic baryons.  

\begin{figure}[h]
 \begin{center}
\begin{tabular}{cc}
\begin{minipage}[cbt]{0.5\textwidth}
{\includegraphics[width=80mm,clip]{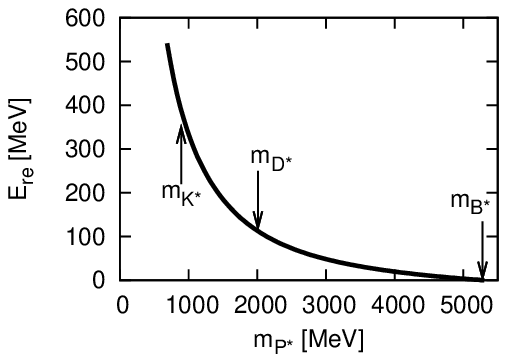}}
\\
\caption{Resonance energy for $(I,J^P)=(0,3/2^-)$ 
as a function of heavy vector meson mass $m_{P^*}$.}
\label{resonancemass}
\end{minipage}
 &\quad\quad
\begin{minipage}[cbt]{0.5\textwidth}
 {\includegraphics[width=80mm,clip]{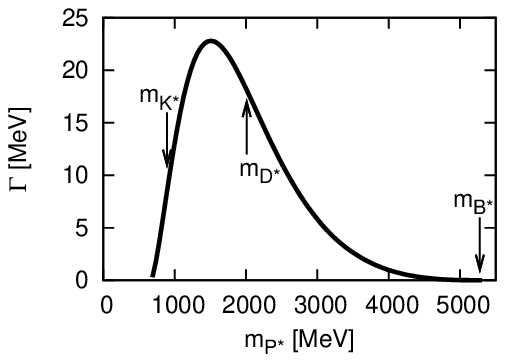}}
\\
\caption{Decay width for $(I,J^P)=(0,3/2^-)$
 as a function of heavy vector meson mass $m_{P^*}$.}
\label{decaymass}
\end{minipage}
\end{tabular}
 \end{center}
\end{figure}

In Fig.~\ref{resonancemass} we plot the resonance energy as a function 
of $m_{P^*}$; it increases as $m_{P^*}$ decreases.  
The resonance exists 
for a small meson mass region also where the bound state 
no longer exists.  
The behavior of the decay width of the resonance as shown in Fig.~\ref{decaymass}
is interesting, as it takes the maximum value at $m_{P^*} \sim 1700$ MeV.  
For larger masses $m_{P^*} \sim 5400$ MeV and beyond, the width becomes 
zero, where the would-be bound state of the single channel 
$P^*N$ is located below the decay channel of $PN$.  
Contrary, as $m_{P^*}$ decreases, the resonance energy becomes 
larger and its wave function extends more.  
This suppresses the overlap of the wave functions of the  
resonance and decaying channels. 
These explain the reason that the decay width takes the maximum value 
at a medium energy point.

\clearpage
\section{Summary}
In this paper, we have investigated heavy baryons having 
exotic flavor quantum numbers as hadronic composites ($\bar{D}N$ and $BN$) of 
a heavy meson and a nucleon.  
In the quark content, this has minimally five quarks of 
$qqqq \bar Q$.  
Because of the light flavor content, the one pion exchange 
interaction is at work and its tensor force plays a very important role
through channel couplings of different angular momentum 
states of $\Delta L = 2$.
The mechanism is essentially the same
as that for the deuteron binding.  
Such channel coupling effects become more striking if 
the masses of the mesons become heavier than those of charmed mesons, 
where the pseudoscalar and vector mesons are more degenerate.  


Our interactions were determined consistently with the low energy
properties of the two-nucleon systems including the deuteron
and some properties of the heavy mesons and the nucleon.
We have confirmed that the previous finding of the pion dominance 
for such novel states to exist by including short range interaction
mediated by $\rho$ and $\omega$.
As a result, we have found that the bound state appears 
for $(I, J^P) = (0, 1/2^-)$, which has been predicted before.  
We have also found a new resonant state for $(I, J^P) = (0, 3/2^-)$ 
with a narrow decay width as a Feshbach resonance predominated by the 
heavy vector meson and nucleon bound state, which decays into an open channel 
of a pseudoscalar and a nucleon.
We have investigated the flavor dependence by changing continuously the masses of heavy mesons.
Then we have found that the bound states in the $(I,J^{P})=(0, 1/2^{-})$
channel exist for the vector meson masses larger than 1700 MeV.
We have also found a resonance with $(I,J^{P})=(0, 3/2^{-})$ having a
narrow decay width.
Therefore the heavy masses of the mesons are crucial for the existence
of the exotic baryons in the present study.


Final remark is related to how these states are observed. 
Resonant states may be found as a pair of 
strongly decaying pseudoscalar meson and nucleon in high energy 
hadronic collisions.  
Exclusive experiments such as hadron and photon induced 
reactions would be useful for the observation of 
bound states since they are stable against strong decay. 
The understanding of production reactions is an important issue
in the future, which can be studied at J-PARC, GSI and so on.
It has been recently proposed that the quark-gluon plasma formed in the relativistic heavy ion collisions will serve a source of multi-particles including exotic hadrons \cite{Cho:2010db}.
Therefore, RHIC, LHC and other facilities will also help to search the exotic states predicted in the present work.

\section*{Acknowledgments}
This work is supported in part by Grant-in-Aid for Scientific 
Research on Priority Areas ¡ÈElucidation of New Hadrons
 with a Variety of Flavors (E01:21105006)¡É from
the ministry of Education, Culture, Sports, Science and Technology
of Japan.

\appendix
\section{Potentials and kinetic terms}
\label{appendix_a}

 Interaction potentials are derived by using the Lagrangians 
Eqs.~\eqref{LpiPP*}-\eqref{LvNN}. In deriving the potentials we use
the static approximation where the energy transfer can be
neglected as compared to the momentum transfer.
The resulting potentials for the coupled channel systems are 
given in the matrix form of $3 \times 3$ for $J^{P}=1/2^-$ and
 of $4 \times 4$ for $J^{P}=3/2^-$,
 
\begin{eqnarray}
 V_{1/2^-} &=
\begin{pmatrix}
V^{11}_{1/2^-} & V^{12}_{1/2^-} & V^{13}_{1/2^-} \\
V^{21}_{1/2^-} & V^{22}_{1/2^-} & V^{23}_{1/2^-} \\
V^{31}_{1/2^-} & V^{32}_{1/2^-} & V^{33}_{1/2^-}
\end{pmatrix}
\, , \\
 V_{3/2^-} &=
\begin{pmatrix}
V^{11}_{3/2^-} & V^{12}_{3/2^-} & V^{13}_{3/2^-} & V^{14}_{3/2^-} \\
V^{21}_{3/2^-} & V^{22}_{3/2^-} & V^{23}_{3/2^-} & V^{24}_{3/2^-} \\
V^{31}_{3/2^-} & V^{32}_{3/2^-} & V^{33}_{3/2^-} & V^{34}_{3/2^-} \\
V^{41}_{3/2^-} & V^{42}_{3/2^-} & V^{43}_{3/2^-} & V^{44}_{3/2^-}
\end{pmatrix}
\, ,
\label{matpotential}
\end{eqnarray}
with the basis given in Table~\ref{table_qnumbers} in the same ordering.
The $\pi$ exchange potential between heavy flavor meson and 
 nucleon is obtained by
\begin{equation}
 V^{\pi}_{1/2^-} = \frac{g_{\pi} g_{\pi NN}}{\sqrt{2}m_N f_{\pi}}
\frac{1}{3} 
\begin{pmatrix}
 0 & \sqrt{3} C_{m_{\pi}} & -\sqrt{6}T_{m_{\pi}}  \\
\sqrt{3}C_{m_{\pi}} & -2C_{m_{\pi}} & -\sqrt{2} T_{m_{\pi}} \\
-\sqrt{6}T_{m_{\pi}} & -\sqrt{2}T_{m_{\pi}} & C_{m_{\pi}} - 2T_{m_{\pi}}
\end{pmatrix}
\vec{\tau}_{P} \cdot \vec{\tau}_N ,
\label{matpi}
\end{equation}

\begin{equation}
 V^{\pi}_{3/2^-} = \frac{g_{\pi}g_{\pi NN}}{\sqrt{2} m_N f_{\pi}}
\frac{1}{3} 
\begin{pmatrix}
 0 & \sqrt{3}T_{m_{\pi}} & -\sqrt{3}T_{m_{\pi}} & \sqrt{3}C_{m_{\pi}} \\
\sqrt{3}T_{m_{\pi}} & C_{m_{\pi}} & 2T_{m_{\pi}} & T_{m_{\pi}} \\
-\sqrt{3}T_{m_{\pi}} & 2T_{m_{\pi}} & C_{m_{\pi}} & -T_{m_{\pi}} \\
\sqrt{3}C_{m_{\pi}} & T_{m_{\pi}} & -T_{m_{\pi}} & -2C_{m_{\pi}}
\end{pmatrix}
\vec{\tau}_{P} \cdot \vec{\tau}_N \, ,
\end{equation}
where $C_m=C(r;m)$,  $T_m=T(r;m)$, and $\vec{\tau_{P}}$ 
and $\vec{\tau}_{N}$ are the isospin matrices for $P(P^{\ast})$ and $N$.
The functions $C(r;m)$ and $T(r;m)$ are given by
\begin{align}
 C(r;m) & = \int \frac{d^3 q}{(2 \pi)^3} \frac{m^2}{\vec{q }\,^2 + m^2}
 e^{i\vec{q}\cdot \vec{r}} F(\Lambda_P,\vec{q}\,)F(\Lambda_N,\vec{q}\,) ,    \label{C} \\ 
T(r;m) S_{12}(\hat{r}) & = \int  \frac{d^3 q}{(2 \pi)^3} 
\frac{-\vec{q}\,^2}{\vec{q}\,^2 + m^2} S_{12}(\hat{q})
 e^{i\vec{q}\cdot \vec{r}} F(\Lambda_P,\vec{q}\,)F(\Lambda_N,\vec{q}\,) ,
\label{T}
\end{align}
with $S_{12}(\hat{x}) = 3(\vec{\sigma}_1 \cdot \hat{x}) (\vec{\sigma}_2
\cdot \hat{x}) -\vec{\sigma}_1 \cdot \vec{\sigma}_2$, and 
$F(\Lambda,\vec{q}\,)$ denotes the form factor given in Eq.~(\ref{eq:cutoff}).

 The corresponding potentials of the $\rho$ meson exchange 
are given by 
\begin{align}
 V^{\rho}_{1/2^-} = & \frac{g_V g_{\rho NN} \beta }{\sqrt{2}m_{\rho}^2}
  \begin{pmatrix}
   C_{m_{\rho}}& 0 & 0 \\
   0 & C_{m_{\rho}} & 0 \\
   0 & 0 & C_{m_{\rho}}
  \end{pmatrix} 
  \vec{\tau}_P \cdot \vec{\tau}_N   \notag \\
 & +\frac{g_V g_{\rho NN} \lambda (1 + \kappa)}{\sqrt{2} m_N}\frac{1}{3}
 \begin{pmatrix}
  0 & 2\sqrt{3} C_{m_{\rho}} & \sqrt{6}T_{m_{\rho}}  \\
  2\sqrt{3}C_{m_{\rho}} & -4C_{m_{\rho}} & \sqrt{2} T_{m_{\rho}} \\
  \sqrt{6}T_{m_{\rho}} & \sqrt{2}T_{m_{\rho}} & 2C_{m_{\rho}} + 2T_{m_{\rho}}
 \end{pmatrix}
\vec{\tau}_{P} \cdot \vec{\tau}_N ,
\label{matrho}
\end{align}
\begin{align}
 V^{\rho}_{3/2^-} = & \frac{g_V g_{\rho NN} \beta }{\sqrt{2}m_{\rho}^2}
  \begin{pmatrix}
   C_{m_{\rho}}& 0 & 0 &0 \\
   0 & C_{m_{\rho}} & 0 &0\\
   0 & 0 & C_{m_{\rho}} & 0 \\
   0 & 0 & 0 & C_{m_{\rho}} 
  \end{pmatrix}
\vec{\tau}_{P} \cdot \vec{\tau}_N \notag \\
& + \frac{g_V g_{\rho NN} \lambda (1 + \kappa)}{ \sqrt{2}m_N}\frac{1}{3}
\begin{pmatrix}
 0 & - \sqrt{3}T_{m_{\rho}} & \sqrt{3}T_{m_{\rho}} & 2\sqrt{3}C_{m_{\rho}} \\
-\sqrt{3}T_{m_{\rho}} & 2C_{m_{\rho}} & -2T_{m_{\rho}} & -T_{m_{\rho}} \\
\sqrt{3}T_{m_{\rho}} & -2T_{m_{\rho}} & 2C_{m_{\rho}} & T_{m_{\rho}} \\
2\sqrt{3}C_{m_{\rho}} & -T_{m_{\rho}} & T_{m_{\rho}} & -4C_{m_{\rho}}
\end{pmatrix}
\vec{\tau}_{P} \cdot \vec{\tau}_N . \notag \\
\label{matrho3/2}
\end{align}
The $\omega$ meson exchange potential can be obtained by 
replacing the relevant coupling constants and the mass of 
the exchanged meson, and by removing the isospin factor 
$\vec{\tau}_{P} \cdot \vec{\tau}_N$.
The anomalous coupling $\kappa$ for the $\omega$ meson exchange
potential is set as zero in Eqs.~\eqref{matrho}-\eqref{matrho3/2}.

\begin{figure}[htbp]
  \begin{center}
   \begin{tabular}{cc}
    $V_{1/2^-}^{11}:\bar{D}N(^2S_{1/2})-\bar{D}N(^2S_{1/2})$&$V_{1/2^-}^{12}:\bar{D}N(^2S_{1/2})-\bar{D}^\ast
    N(^2S_{1/2})$ \\
 \includegraphics[width=80mm,clip]{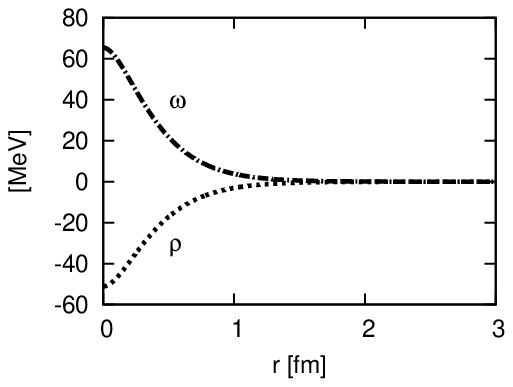} &
 \includegraphics[width=80mm,clip]{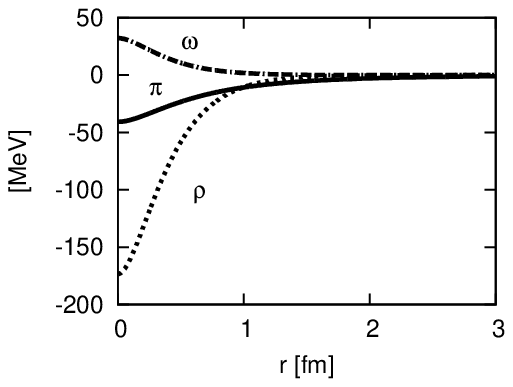} \\[2mm]
    $V_{1/2^-}^{13}:\bar{D}N(^2S_{1/2})-\bar{D}^\ast N(^4D_{1/2})$&$V_{1/2^-}^{22}:\bar{D}^\ast N(^2S_{1/2})-\bar{D}^\ast
    N(^2S_{1/2})$ \\
 \includegraphics[width=80mm,clip]{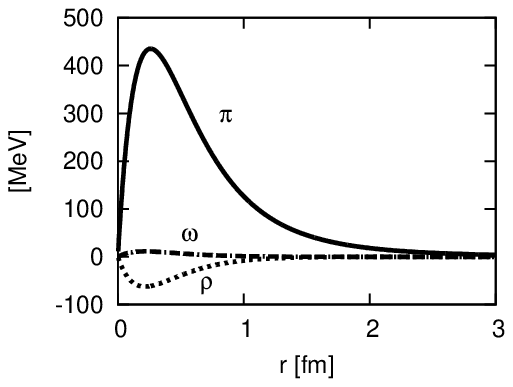} &
 \includegraphics[width=80mm,clip]{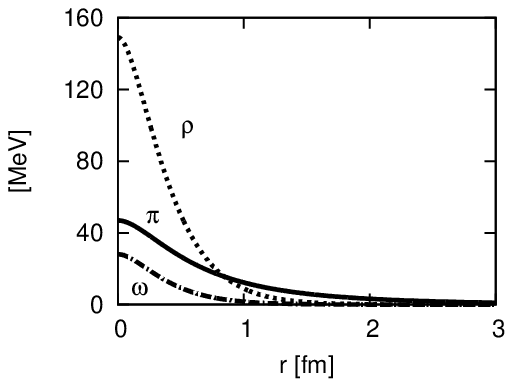} \\[2mm]
    $V_{1/2^-}^{23}:\bar{D}^\ast N(^2S_{1/2})-\bar{D}^\ast N(^4D_{1/2})$&$V_{1/2^-}^{33}:\bar{D}^\ast N(^4D_{1/2})-\bar{D}^\ast
    N(^4D_{1/2})$ \\
 \includegraphics[width=80mm,clip]{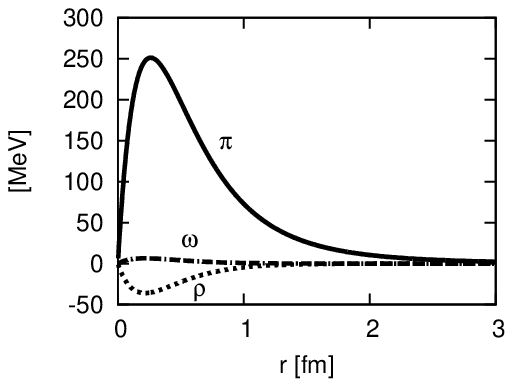} &
 \includegraphics[width=80mm,clip]{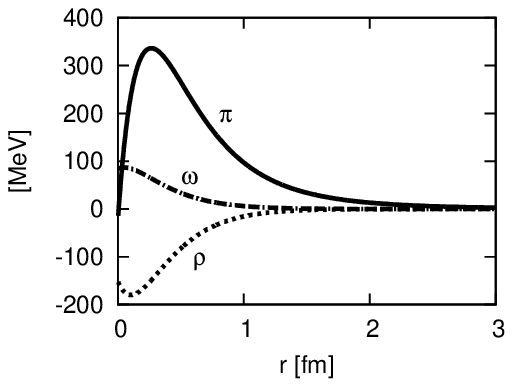} \\
   \end{tabular}
   \caption{Various components of the $\pi\rho\,\omega$ exchange potential for
   $(I,J^P)=(0,1/2^-)$.}
\label{pote1/2}
  \end{center}
 \end{figure}

In Figs.~\ref{pote1/2}-\ref{pote3/2b} the functional forms of all the potentials are shown.
In these figures, we can see clearly the dominance of 
the tensor force in the transition amplitude, for instance, 
$V^{13}_{1/2^{-}}$ and $V^{23}_{1/2^{-}}$.

The kinetic terms are given by
\begin{align}
 K_{1/2^-} &= \mbox{diag} \left( -\frac{1}{2 \tilde{m}_P}\bigtriangleup_0 ,
- \frac{1}{2\tilde{m}_{P^*}} \bigtriangleup_0 +\Delta m_{PP^*},
-\frac{1}{2\tilde{m}_{P^*}} \bigtriangleup_2 +\Delta m_{PP^*} \right)\,
 , \\
K_{3/2^-} &= \mbox{diag} \left( -\frac{1}{2 \tilde{m}_P}\bigtriangleup_2 ,
- \frac{1}{2\tilde{m}_{P^*}} \bigtriangleup_0 +\Delta m_{PP^*},
-\frac{1}{2\tilde{m}_{P^*}} \bigtriangleup_2 +\Delta m_{PP^*}
\right. , \notag \\
& \quad \left.-\frac{1}{2\tilde{m}_{P^*}} \bigtriangleup_2 +\Delta
 m_{PP^*} \right) \, ,
\end{align}
for $J^P = 1/2^-$ and $3/2^-$, respectively. Here, we define
$\bigtriangleup_0 = \partial^2 / \partial r^2 + (2/r)\partial / \partial
r$
and $\bigtriangleup_2 = \bigtriangleup_0 +6/r^2 , \, \tilde{m}_{P{^{(\ast)}}}
= m_N m_{P{^{(\ast)}}}/(m_N +m_{P{^{(\ast)}}}),$ with $\Delta m_{PP^*} = m_{P^*} -m_P$.
The total Hamiltonian is then given by $H_{J^P} = K_{J^P} + V_{J^P}$.

 \begin{figure}[htbp]
  \begin{center}
   \begin{tabular}{cc}
    $V_{3/2^-}^{11}:\bar{D}N(^2D_{3/2})-\bar{D}N(^2D_{3/2})$&$V_{3/2^-}^{12}:\bar{D}N(^2D_{3/2})-\bar{D}^\ast
    N(^4S_{3/2})$ \\
 \includegraphics[width=80mm,clip]{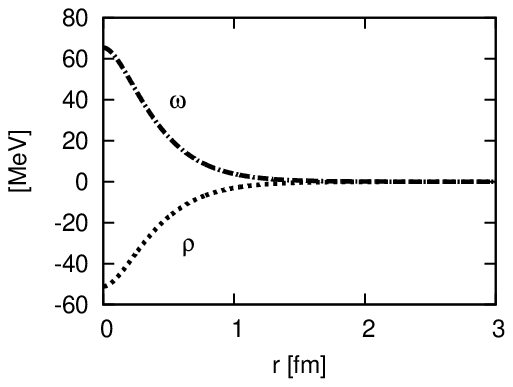} &
 \includegraphics[width=80mm,clip]{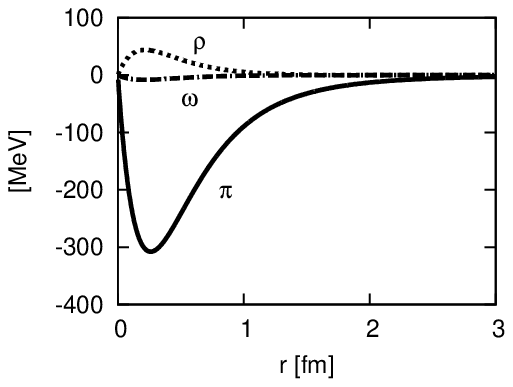} \\[2mm]
    $V_{3/2^-}^{13}:\bar{D}N(^2D_{3/2})-\bar{D}^\ast N(^4D_{3/2})$&$V_{3/2^-}^{14}:\bar{D}N(^2D_{3/2})-\bar{D}^\ast
    N(^2D_{3/2})$ \\
 \includegraphics[width=80mm,clip]{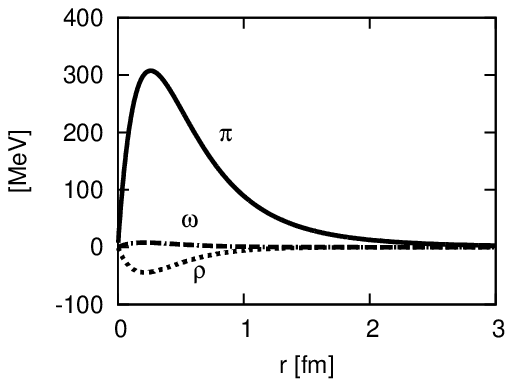} &
 \includegraphics[width=80mm,clip]{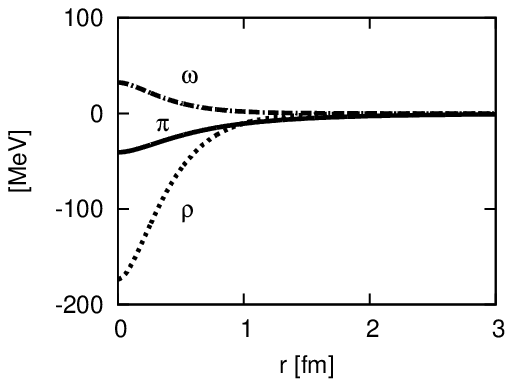} \\[2mm]
   \end{tabular}
\caption{Various components of the $\pi\rho\,\omega$ exchange potential 
for $(I,J^P)=(0,3/2^-)$.}
\label{pote3/2a}
  \end{center}
 \end{figure}

\begin{figure}[htbp]
 \begin{center}
  \begin{tabular}{cc}
    $V_{3/2^-}^{22}:\bar{D}^\ast N(^4S_{3/2})-\bar{D}^\ast N(^4S_{3/2})$&$V_{3/2^-}^{23}:\bar{D}^\ast N(^4S_{3/2})-\bar{D}^\ast
    N(^4D_{3/2})$ \\
 \includegraphics[width=80mm,clip]{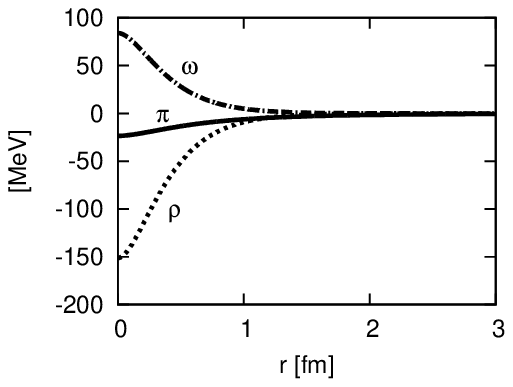} &
 \includegraphics[width=80mm,clip]{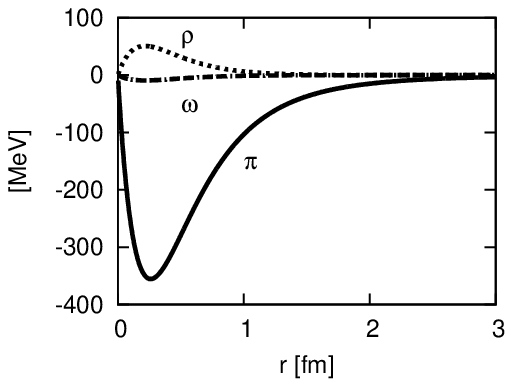} \\[2mm]
    $V_{3/2^-}^{24}:\bar{D}^\ast N(^4S_{3/2})-\bar{D}^\ast N(^2D_{3/2})$&$V_{3/2^-}^{33}:\bar{D}^\ast N(^4D_{3/2})-\bar{D}^\ast
    N(^4D_{3/2})$ \\
 \includegraphics[width=80mm,clip]{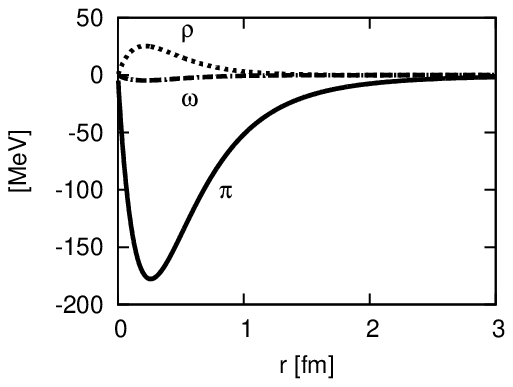} &
 \includegraphics[width=80mm,clip]{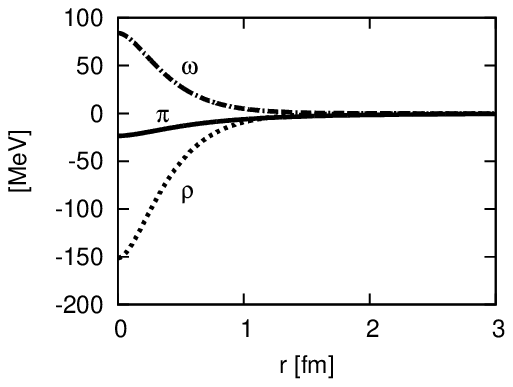} \\[2mm]
    $V_{3/2^-}^{34}:\bar{D}^\ast N(^4D_{3/2})-\bar{D}^\ast N(^2D_{3/2})$&$V_{3/2^-}^{44}:\bar{D}^\ast N(^2D_{3/2})-\bar{D}^\ast
    N(^2D_{3/2})$ \\
 \includegraphics[width=80mm,clip]{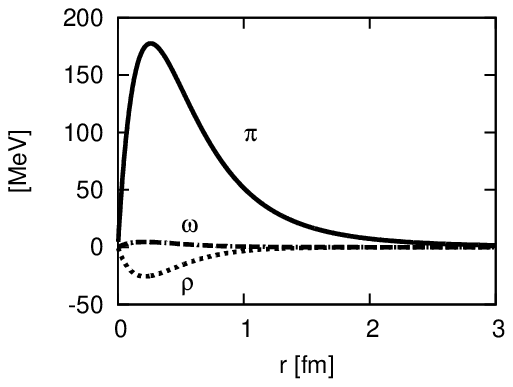} &
 \includegraphics[width=80mm,clip]{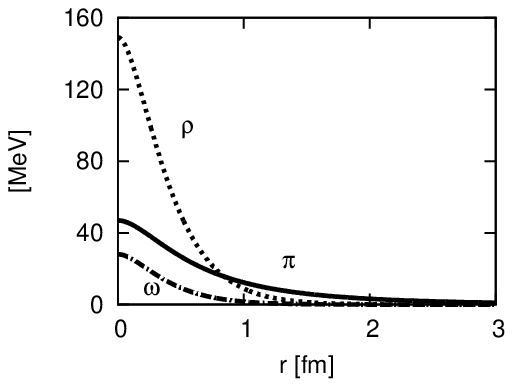} \\
  \end{tabular}
\caption{Continued from Fig.~\ref{pote3/2a}.}
\label{pote3/2b}
 \end{center}
\end{figure}
\clearpage


\end{document}